\documentclass[letter,nobibnotes,nofootinbib]{revtex4}

\usepackage{amsmath,amssymb}
\usepackage{epsfig}
\usepackage{graphicx}
\usepackage{xspace}

\newcommand{\be}{\begin{equation}}
\newcommand{\ee}{\end{equation}}
\newcommand{\bL}{\begin{Large}}
\newcommand{\eL}{\end{Large}}
\newcommand{\ba}{\begin{eqnarray}}
\newcommand{\ea}{\end{eqnarray}}
\newcommand{\bc}{\begin{center}}
\newcommand{\ec}{\end{center}}
\newcommand{\bfig}{\begin{figure}}
\newcommand{\efig}{\end{figure}}
\newcommand{\f}[2]{\frac{#1}{#2}}
\newcommand{\g}{\gamma}
\newcommand{\om}{\omega}

\newcommand{\la}{\label}

\newcommand{\sg}{\sigma}

\newcommand{\al}{\alpha}

\newcommand{\rr}[4]{#1, {\it #2 \/}{\bf #3} #4}

\newcommand{\mW}{\ensuremath{m_{\mathrm{W}}}\xspace}
\newcommand{\mWfit}{\ensuremath{m_{\mathrm{W}}^{\mathrm{fit}}}\xspace}
\newcommand{\mWmin}{\ensuremath{m_{\mathrm{W}}^{\mathrm{min}}}\xspace}

\newcommand{\ff}{\ensuremath{{\cal F}}\xspace}

\newcommand{\GeV}{\ensuremath{\mbox{GeV}}\xspace}

\newcommand{\Ifb}{\ensuremath{\mathrm{fb^{-1}}}\xspace}
\newcommand{\ttbar}{\ensuremath{\mathrm{t\bar{t}}}\xspace}
\newcommand{\WW}{\ensuremath{\mathrm{W^+W^-}}\xspace}
\newcommand{\W}{\ensuremath{\mathrm{W}}\xspace}

\begin{document}
\title{Threshold scans in diffractive W pair production via QED
processes at the LHC}

\author{M. Boonekamp}\email{boon@hep.saclay.cea.fr} 
\affiliation{Service de physique des particules, CEA/Saclay,
  91191 Gif-sur-Yvette cedex, France}
\author{J. Cammin}\email{cammin@fnal.gov} 
\affiliation{University of Rochester, New York, USA}
\author{R. Peschanski}\email{pesch@spht.saclay.cea.fr}
\affiliation{Service de physique th{\'e}orique, CEA/Saclay,
  91191 Gif-sur-Yvette cedex, France\footnote{%
URA 2306, unit{\'e} de recherche associ{\'e}e au CNRS.}}
\author{C. Royon}\email{royon@hep.saclay.cea.fr}
\affiliation{Service de physique des particules, CEA/Saclay,
  91191 Gif-sur-Yvette cedex, France}

\begin{abstract}
We propose a new set of measurements which can be performed at
the LHC using roman pot detectors.
This new method is based on exploiting excitation curves to measure kinematical
properties of produced particles. We illustrate it in the case of central 
diffractive $W$ pair production.
\end{abstract}

\maketitle

\section{Introduction}

We propose a new method to measure heavy particle properties via double 
photon exchange at the LHC. In this category of events, the heavy objects 
are produced in pairs, whereas the beam particles
often leave the interaction region intact, and can be measured using very forward detectors.

If the events are $exclusive$, \emph{i.e.}, if no other particles are produced in addition to the pair of heavy objects 
and the outgoing protons, the proton measurement gives access to the photon-photon 
centre-of-mass, and the dynamics of the hard process can be accurately studied. 
In particular, one can observe the threshold excitation and attempt to extract the mass of the heavy particle, 
or study its (possibly energy-dependent) couplings by measuring cross-sections and angular 
distributions \cite{piotr}. As examples of this approach, we give a detailed account of the \W boson 
measurement at production threshold. The method can easily be extended to other heavy objects
in exclusive production.

The Letter is organised as follows. We start by giving the theoretical 
formulation  of \WW production (via QED). We then describe the event generation,
the simulation of detector effects, and the cuts used in the analysis.
The following part of the paper describes in detail the threshold scan method, 
in a twofold version (``turn-on''and ``histogram'' fits), and its application to 
the \W boson measurements.

\section{Theoretical formulation of $W$ pair QED production}

The QED process rates are obtained from  the following  cross section formula
\be
d\sg_{(\mathrm{pp\to \ p\ W^+ W^-p})} = \hat {\sg}_{\mathrm{\g\g\to W^+ W^-}}\ dn^{\g}_1 \ 
dn^{\g}_2\nonumber \ ,
\la{dsigmaWWW}
\ee
\
where the Born $\mathrm{\g\g\to W^+ W^-}$ cross-section reads 
\cite{Papageorgiu:1990mu}
\be
\hat {\sg}_{\mathrm{\g\g\to W^+ W^-}}=\frac {8\pi\al^2}{M_{\mathrm{WW}}^2}\left\{
\f 1t\left(1+\f34t+3t^2\right)\Lambda-3t(1-2t)
\ln\left(\f{1+\Lambda}{1-\Lambda}\right)\right\},
\la{hatsigma}
\ee
with
\be
t= \f{m_\W^2}{M_{\mathrm{WW}}^2}\ ,\ \ \ \ \ \ \ \ \ \ \Lambda =\sqrt{1-4t}\ , 
\la{kine}
\ee
where $M_{\mathrm{WW}}$ is the total \WW mass. The photon fluxes $dn^{\g}$ are given 
by  
\cite{Budnev:1975zs}
\be
 dn^{\g}=\f{\al}{\pi} \f{\om}{\om}\left(1-\f{\om}{E}\right)
\left[\phi\left(\f{q^2_{max}}{q^2_{0}}\right)-\phi\left(\f{q^2_{min}}{q^2_{0}
}
\right)\right]\ ,
\ee
where
\be
\phi\left(x\right)\equiv (1+ay)
\left[\ln\left(\f x{1+x}\right)+\Sigma_{k=1}^3 \f1{k(1+x)^k}\right]
-\ \f{(1-b)y}{4x(1+x)^3+c(1+\f14 y)}
\left[\ln\left(\f {2+2x-b}{1+x}\right)+\Sigma_{k=1}^3 
\f{b^k}{k(1+x)^k}\right]\ ,
\la{flux}
\ee
and
\be
q^2_{0} \sim 0.71\ \mathrm{GeV}^2\ \ ;\ y= \f{\om^2}{E(E-\om)}\ \ ;\ a \sim 7.16\ \ ;\ 
b\ \  
\sim -3.96\ \ ;\ c\sim 0.028 \ , 
\la{kineflux}
\ee
in the usual dipole approximation for the proton electromagnetic form 
factors. $\om$ 
is the photon energy in the laboratory frame, $q^2$ the modulus of its mass 
squared in 
the range
\be
\left[q^2_{min},\ q^2_{max}\right]\equiv \left[\f{m^2\om^2}{E(E-\om)}\ , 
\f{t_{max}}{q^2_{0}}\right]\ ,
\la{range}
\ee  
where $E$ and $m$ are  the energy and mass of the incident particle and  
$t_{max}\equiv ({m_\W^2}/{M_{\W\W}^2})_{max} $ is defined by the experimental 
conditions.

The QED cross section $d\sg (\mathrm{pp\to \ p\ W^+ W^-p})$ is  a theoretically 
clear 
prediction. One should take into account however, two sources of 
correction factors. One is due to the soft QCD initial state radiation 
between 
incident protons which could destroy the large rapidity gap of the QED 
process. It 
is present but much less pronounced than for the rapidity gap survival for a 
QCD 
hard process (see the  discussion in the next subsections), thanks to 
the large 
impact parameter implied by the QED scattering.  
The second factor   is the QCD $\mathrm{gg \to  W^+ W^-}$ exclusive production via 
higher 
order diagrams. This remains to be evaluated. In standard recently (non diffractive) production
\cite{Binoth:2005ua}, it is small. The similar
calculation for 
the diffractive \WW production by comparison with the QED process is outside 
the 
scope of our paper but deserves to be studied together with the ``inclusive'' 
background (\WW{}+hadrons) it could generate.

\subsection{Rapidity Gap Survival}

In order to select exclusive diffractive states, such as for \WW (QED), 
it is required to take into account the 
corrections from soft hadronic scattering. Indeed, the soft scattering  
between incident particles tends to mask the genuine
hard diffractive interactions at 
hadronic colliders. Starting with the ``hard" scattering amplitude ${\cal A}_{(\mathrm{WW})}$, 
the formulation of this correction \cite{sp,pom} consists
in considering its convolution with a soft S-matrix elemeny $S$ which reflects
the small ``rapidity gap survival" factor due to the soft radiation always
present when two initial hadrons collide \cite{sp}. One writes
\begin{equation}
{\cal A}(p_{T1},p_{T2}, \Delta \Phi) =
\left\{ 1 +{\cal A}_{SP} \right\}{\bf \times} {\cal A}_{(\mathrm{WW},\ttbar)}\equiv {\cal S} 
{\bf \times} 
{\cal A}_{(\mathrm{WW},\ttbar)} = \int d^2{\bf k}_T\ {\cal S}({\bf k}_T) \ {\cal 
A}_{(\mathrm{WW},\ttbar)}({\bf p}_{T1}\!-\!{\bf k}_T,
{\bf p}_{T2}\!+
\!{\bf k}_T) 
\ ,  
\label{sp}
\end{equation}
where ${\bf p}_{T1,2}$ are the transverse momenta of the outgoing $p,\bar p$ 
and $\Delta \Phi$ their 
azimuthal angle separation.  
${\bf k}_T$ is the intermediate transverse momentum integrated out by the convolution.

The correction for the QED process 
is present but much less pronounced than for the rapidity gap survival for a 
QCD 
hard process, thanks to 
the large 
impact parameter implied by the QED scattering. In a specific model 
\cite{Khoze:2001xm} the correction factor  has 
been 
evaluated to be of order $0.9$ at the LHC for $\g\g\to \mathrm{H}$
and by contrast, $0.03$ for the QCD exclusive diffractive 
processes at the LHC.

\section{Experimental context}

\subsection{The DPEMC Monte Carlo}
A recently developed Monte-Carlo program, {\tt DPEMC} \cite{dpemc}, provides 
an implementation of the \WW  events described above in the
QED  exchange modes.
It uses {\tt HERWIG} \cite{herwig} as a cross-section library of
hard QCD 
processes, and when required, convolutes them with the relevant pomeron 
densities. HERWIG is only used for parton sjowering and hadronisation
for exclusive processes.
The survival probabilities discussed in the 
previous section 
(0.9 for double photon exchange 
processes)
have been introduced at generator level. The cross section at generator
level for \WW QED  is found to be 55.9 fb for a $m_\W$ mass of 
80.42
GeV after applying the survival probabilities.

\subsection{Roman pot detector positions and resolutions}

A possible experimental setup for forward proton detection is described in 
detail in
\cite{helsinki}. We will only describe its main features here and discuss its
relevance for the \W boson and top quark masses measurements.

In exclusive QED processes, 
the mass of the central heavy object can be reconstructed
using the roman pot detectors and tagging both protons
in the final state at the LHC. It is given  by $M^2 = \xi_1\xi_2 s$, where 
$\xi_i$ are 
the proton fractional momentum losses, and $s$  the total center-of-mass 
energy squared \cite{Albrow:2000na}. In order to reconstruct objects with masses in the 160 GeV
range (for \WW events) in this 
way, the acceptance should be large down to $\xi$ values as low as a few 
$10^{-3}$. 
The missing mass resolution directly depends on the resolution on 
$\xi$, and should not exceed a few percent to obtain a good mass resolution.

These goals can be achieved if one assumes two detector stations, located at 
$\sim 
220$ m, and $\sim 420$ m \cite{helsinki} from the interaction point. The $\xi$ acceptance and 
resolution have been derived for each device using a complete simulation
of the LHC beam parameters. The combined $\xi$ acceptance is close to $\sim 60\%
$ at low masses (at about twice $m_\W$).   

Our analysis does not assume any particular value for the $\xi$ resolution. 
We will discuss in the following how the resolution on the \W boson 
mass depends on the detector resolutions, or in other words,
the missing mass resolution.

\subsection{Experimental cuts}

Let us summarise the cuts applied in the remaining part of the analysis.
As said before, both diffracted protons are required to be detected in 
roman pot detectors.

The triggers which will be used for the \WW  events will be the
usual ones at the LHC requiring in addition a positive tagging in the roman
pot detectors.

The experimental offline cuts and their efficiencies have been obtained using a
fast simulation of the CMS detector \cite{cmsim} as an example, the fast
simulation of the ATLAS detector \cite{cmsim} leading to the same results.
If we require at least one lepton (electron or muon) with a transverse
momentum greater than 20 GeV and one  jet with a transverse
momentum greater than 20 GeV for \WW  to be
reconstructed in the acceptance of the main detector in addition to the tagged 
protons,
we get an efficiency of about 30\%
for \WW events. We give the mass resolution as a function of luminosity in the
following after taking into account these efficiencies. If the efficiencies are
found to be higher, the luminosities have to be rescaled by this amount.

\section{Threshold scan methods}

\subsection{Explanation of the histogram and turn-on fit methods}
We study two different methods to reconstruct the mass of heavy objects
double diffractively produced at the LHC. As we mentioned before, the method is
based on a fit to the turn-on point of the missing mass distribution at 
threshold. 

One proposed method (the ``histogram'' method) corresponds to the comparison of 
the mass distribution in data with some reference distributions following
 a Monte Carlo simulation of the detector with different input masses
corresponding to the data luminosity. As an example, we can produce 
a data sample for 100 fb$^{-1}$ with different $W$ masses.  
For each Monte Carlo sample, a $\chi^2$ value corresponding to the 
population difference in each bin between data and MC is computed. The mass point 
where
the $\chi^2$ is minimum corresponds to the mass of the produced object in data.
This method has the advantage of being easy but requires a good
simulation of the detector.

The other proposed method (the ``turn-on fit'' method) is less sensitive to the MC 
simulation of the
detectors. The threshold scan is directly sensitive to
the mass of the diffractively produced object (in the \WW case for instance, it
is sensitive to twice the \W mass). The idea is thus to fit the turn-on
point of the missing mass distribution which leads directly to the mass 
of the produced object, the \W boson. Due to its robustness,
this method is considered as the ``default" one in the following.

To illustrate the principle of these methods and their achievements,
we  apply them to the 
\W boson  in the
following, and present in detail the reaches at the LHC. They can be applied to other 
threshold scans as well.

\subsection{W mass measurement using diffractive QED events}

\begin{figure}
  \centering
  \includegraphics[width=0.4\linewidth]{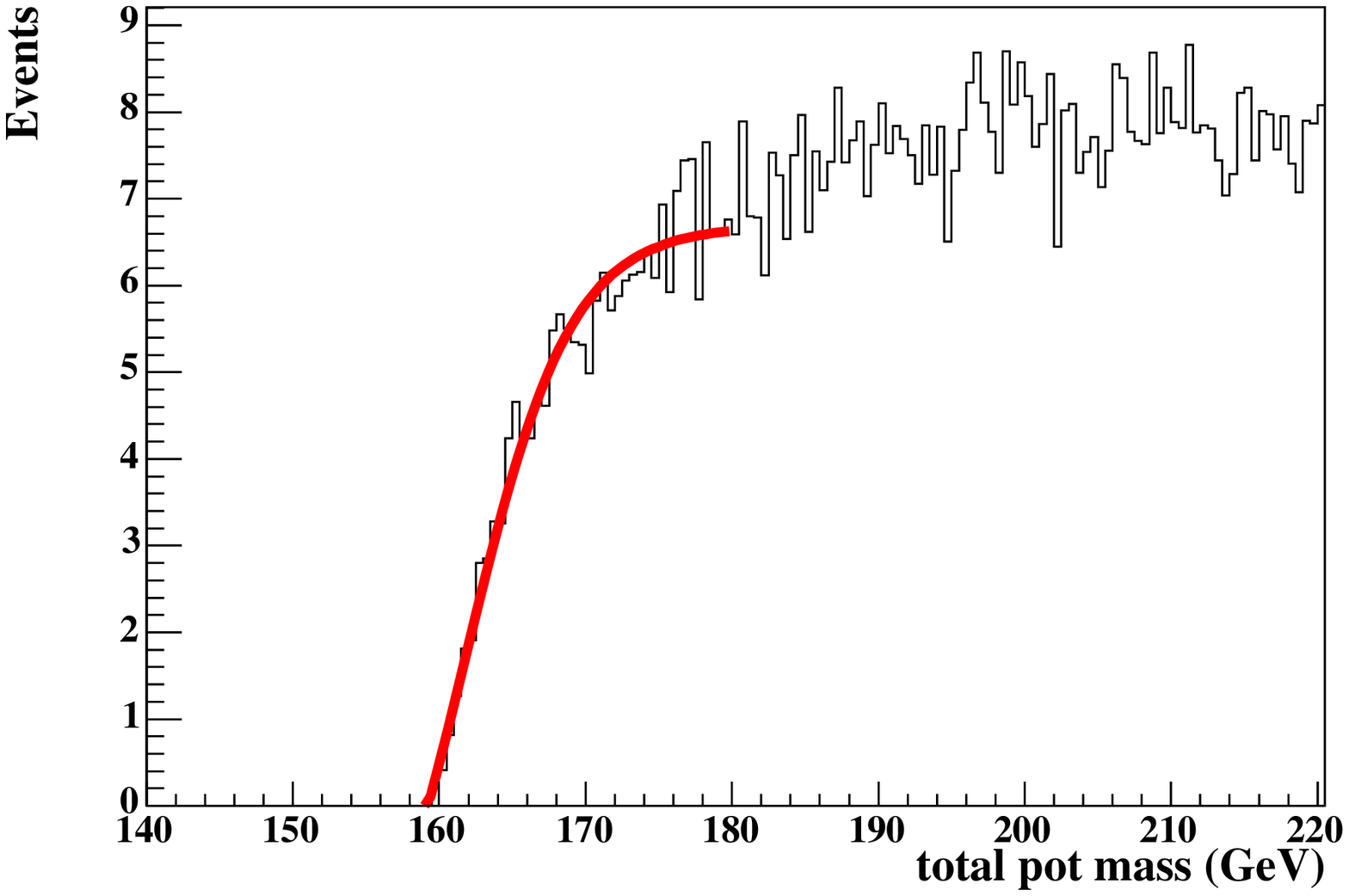}
  \includegraphics[width=0.4\linewidth]{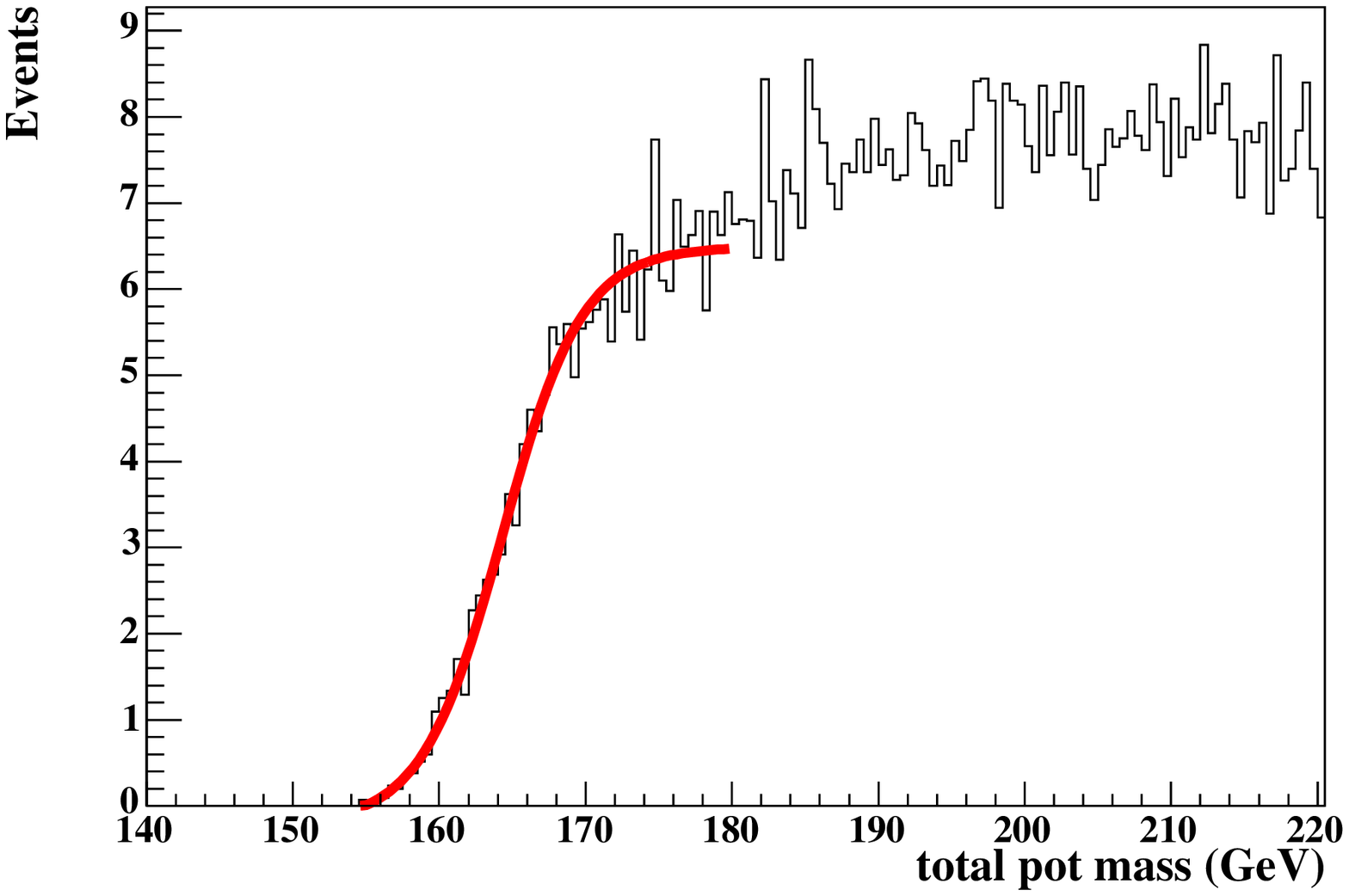}
  \caption{Two examples of fits to missing mass reference distributions with a
  resolution of the roman pot detectors of 1~\GeV (left) and 3~\GeV
  (right). We see on these plots the principle and the accuracy of the ``turn-on 
fits"
  to the MC at threshold. (Please note that the produced events were
  reweighted to a luminosity of 100 fb$^{-1}$ in a standard way explaining
  why the statistical fluctuations are small.)}
  \label{fig:WW_ref_fits}
\end{figure}

\begin{figure}
  \centering
  \includegraphics[width=0.4\linewidth]{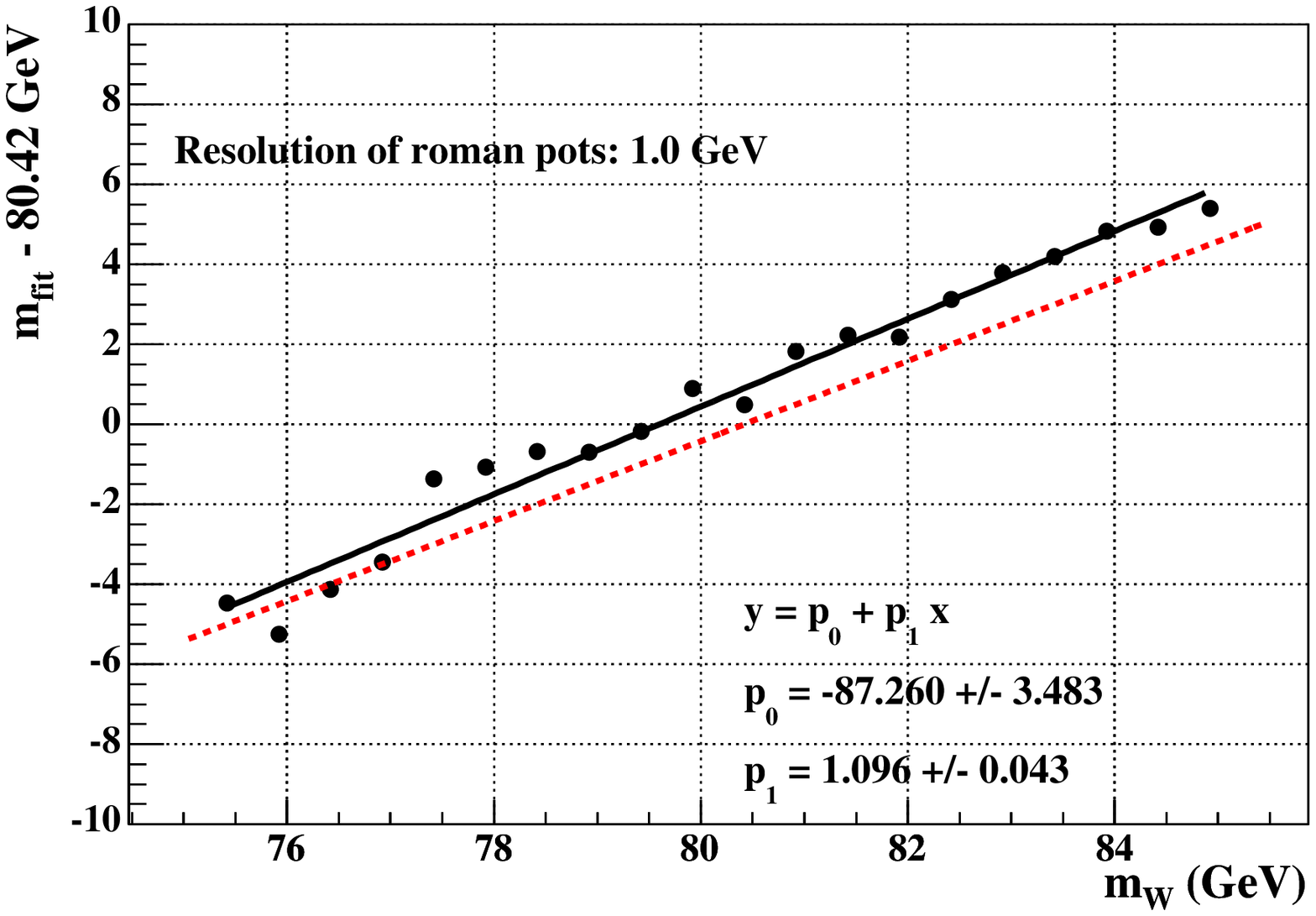}
  \includegraphics[width=0.4\linewidth]{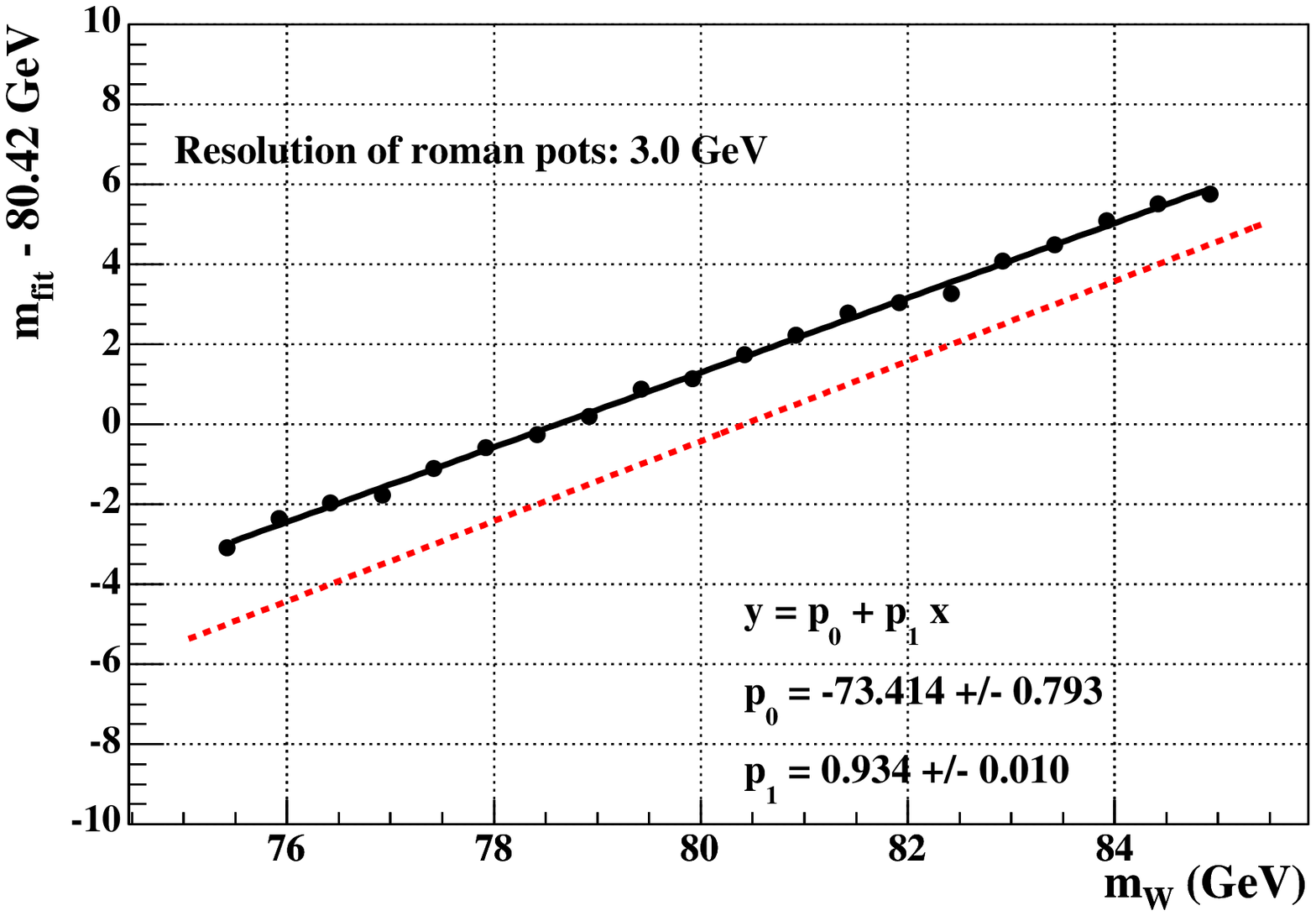}
  \caption{Calibration curves (see text) for two different roman pot resolutions 
of 1~\GeV
  (left) and 3~\GeV (right). We notice that the calibration can be fitted
  to a linear function with good accuracy. The dashed line indicates 
  the first diagonal to show the shift clearly.}
  \label{fig:calibration_WW}
\end{figure}

In this section, we will first describe the result of the ``turn-on fit" 
method to perform a measurement of the \W mass using diffractive QED events.
The advantage
of the \WW processes is that they do not suffer from any theoretical uncertainties
since this is a QED process.
The W mass can be extracted by fitting a 4-parameter `turn-on' curve to the 
threshold
of the mass distribution (c.f. Ref.~\cite{Abbiendi:2002ay}):
\begin{equation}\label{eq:fitfunc}
\ff =   P_1\cdot \left( \left[{e^{-\frac{x-P_2}{P_3}}+1}\right]^{-1}+P_4\right).
\end{equation}
$P_1$ is the amplitude, $P_2$ the inflexion point, $P_3$ the width of
the turn-on curve, and $P_4$ is a vertical offset, $x$ being
the missing mass. With a detector of
perfect resolution, $P_2$ would be equal to twice the \W mass.
However, the finite roman pot resolution leads to
a shift between $P_2$ and $2 m_\W$ which has to be established using a
MC simulation of the detector for different values of its resolution.
This shift is only related to the method itself and does not correspond
to any error in data. For each value of the \W input mass in MC, one has to 
obtain the
shift between the reconstructed mass ($P_2/2$) and the input mass, which
we call in the following the calibration curve.
It is assumed for simplicity that $P_2$ is a linear function of \mW,
which is a good approximation as we will see next. 
In order to determine the linear
dependence between $P_2$ and \mW, calibration curves are calculated
for several assumed resolutions of the roman pot detectors. The
calibration points are obtained by fitting \ff to the mass
distribution of high statistics samples (100\,000 events) for several
values of \mW. An example is given in Fig.~\ref{fig:WW_ref_fits} for
two resolutions of the roman pot detectors. The difference
between the fitted values of $P_2/2$ and the input \W masses
are plotted as a function of the input W mass and are then fitted with a
linear function.  To minimise the errors on the slope and offset, the difference 
$P_2/2-80.42~\GeV$ is plotted versus $\mW$
(Fig.~\ref{fig:calibration_WW}).

To evaluate the statistical uncertainty due to the method itself,
we perform the fits with some 100 different ``data" ensembles.
For each ensemble, one obtains a different 
reconstructed \W mass, the dispersion corresponding only to statistical
effects.
The expected statistical uncertainty on the actual measurement of the
W mass in data is thus estimated with these ensemble tests for several 
integrated
luminosities and roman pot resolutions. Each ensemble contains a
number of events that corresponds to the expected event yield for a
given integrated luminosity, taking into account selection and
acceptance efficiencies. The turn-on function \ff is fitted to each
ensemble. Only the parameters $P_1$ and $P_2$ are allowed to float,
$P_3$ and $P_4$ are fixed to the average values obtained from the fits
for the calibration points.

\begin{figure}
  \centering
  \includegraphics[width=0.35\linewidth]{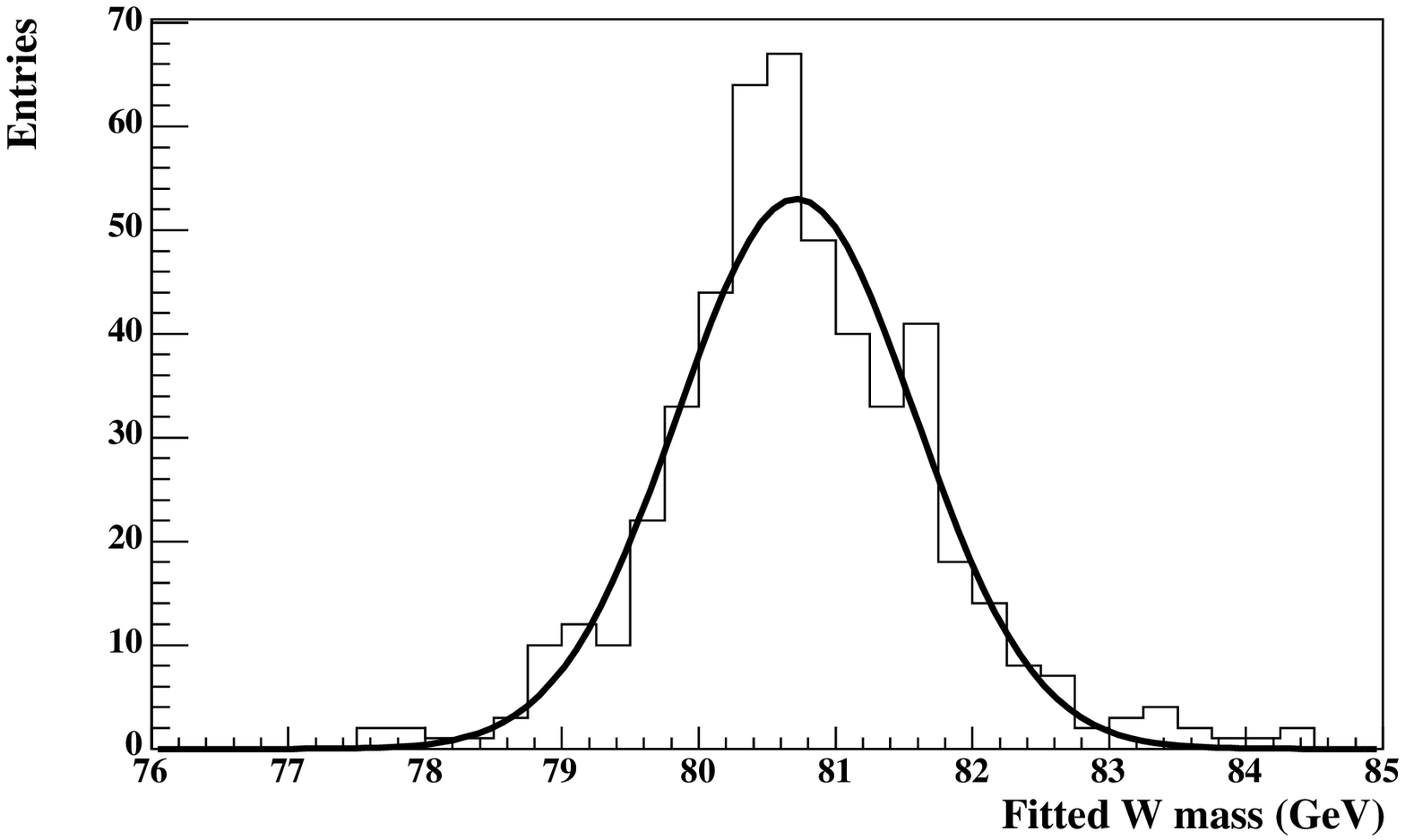}
  \qquad
  \includegraphics[width=0.35\linewidth]{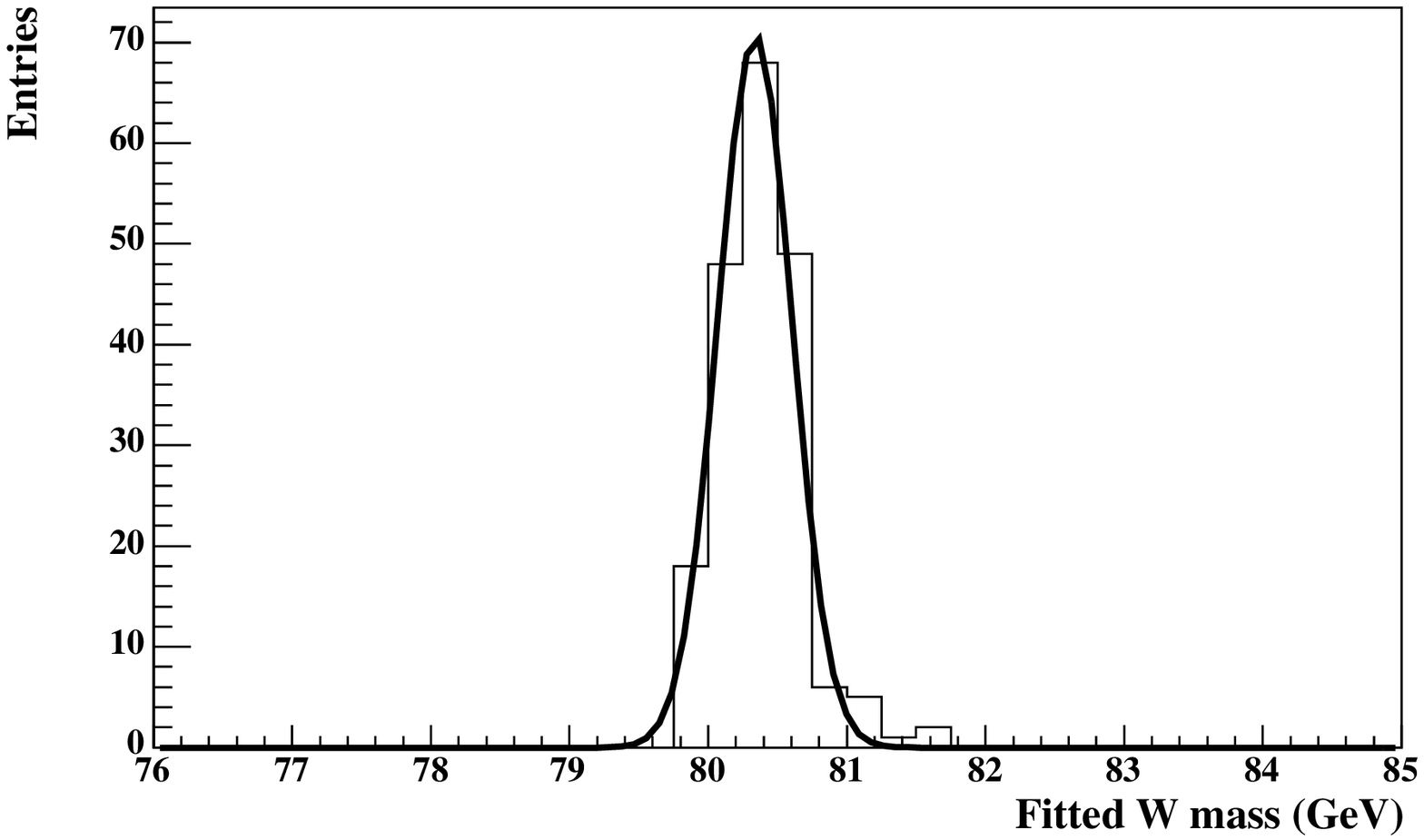}
  \caption{Distribution of the fitted value of the W mass from
  ensemble tests. Left: corresponding to 150~\Ifb , right:
  corresponding to 300~\Ifb.
  We note the resolution obtained on the \W mass for these two luminosities.}
  \label{fig:WW_ensemble_distributions}
\end{figure}

\begin{figure}
  \centering
  \includegraphics[width=0.35\linewidth]{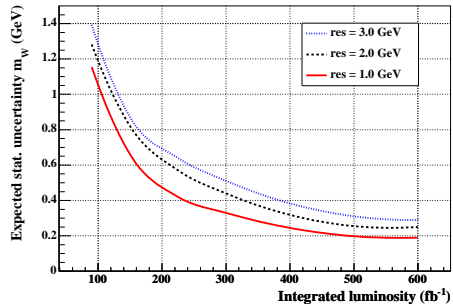}
  \caption{Expected statistical uncertainty on the \W mass
  as a function of luminosity for three different roman
  pot resolutions using the turn-on fit method.
  }
  \label{fig:ww_resvslumib}
\end{figure}

In order to obtain the fitted estimate for the W mass, \mWfit, in each
ensemble, the fit value of $P_2$ is corrected with the calibration
curve that corresponds to the roman pot resolution. For each
resolution \mWfit is histogrammed as shown in
Fig.~\ref{fig:WW_ensemble_distributions}.  The distributions are
fitted with a Gauss function where the width corresponds to the
expected statistical uncertainty of the W mass measurement.
Fig.~\ref{fig:ww_resvslumib} shows the expected precision as a function
of the integrated luminosity for several roman pot resolutions.
With 150~\Ifb the expected statistical uncertainty on \mW is about
0.65~\GeV when a resolution of the roman pot detectors of 1~\GeV can be
reached. With 300~\Ifb the expected uncertainty on \mW decreases to about 0.3~\GeV.

We notice of course that this method is not competitive to get a precise
measurement of the \W mass, which would require a resolution to be better
than 30 MeV. However, this method can be used to align precisely the roman
pot detectors for further measurements. A precision of 1 GeV (0.3 GeV)
on the \W mass leads directly to a relative resolution of 1.2\%
(0.4\%) on $\xi$ using the missing mass method. 
   
\begin{figure}
  \includegraphics[width=0.3\linewidth]{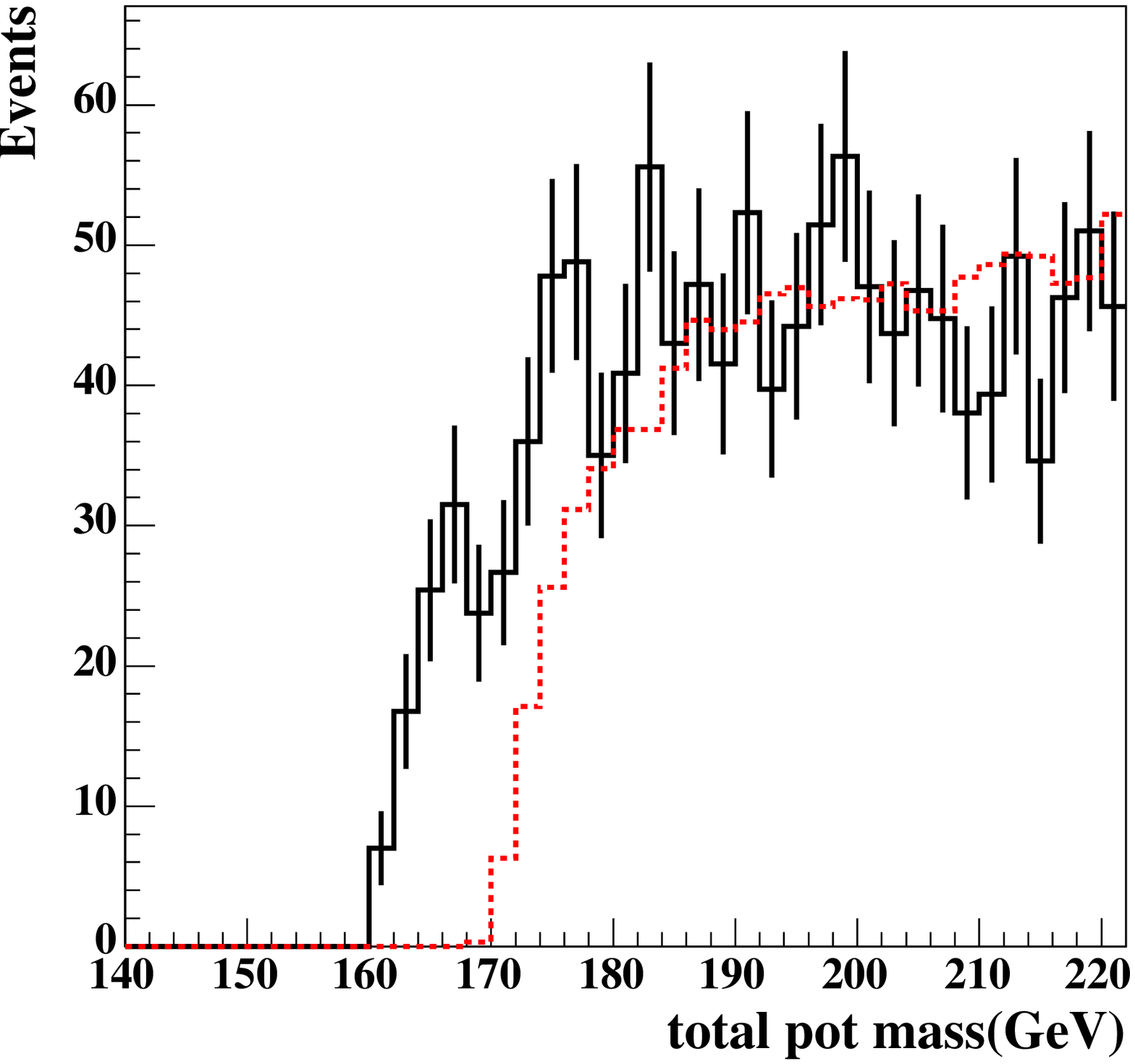}
  \qquad
  \includegraphics[width=0.3\linewidth]{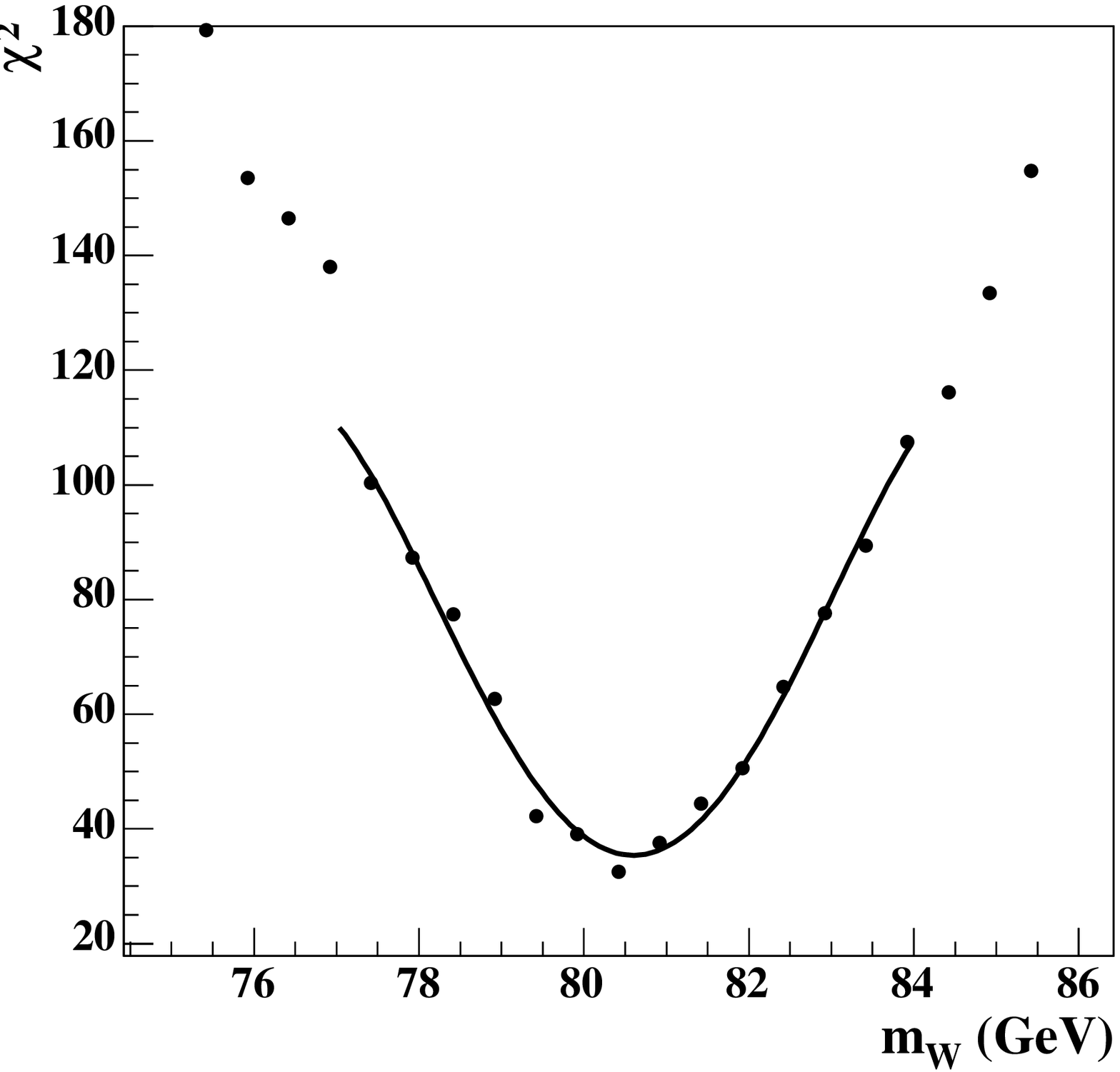}
  \caption{Left: Example of the histogram-fitting method. We see the difference
  between the ``data" sample (full histogram with error bars,
  $\mW=80.42~\GeV$) and a reference histogram (dashed line, $\mW=85.42~\GeV$). 
  Right:
    Example of the $\chi^2$ distribution in one ensemble.}
  \label{fig:W_mass_hist_fit}
\end{figure}

Let us now present the result on the ``histogram" method, which is an
alternative approach to determine the W mass.
The same high statistics templates used to derive the calibration
curves are fitted directly to each ensemble
(see Fig.~\ref{fig:W_mass_hist_fit} left). 
The $\chi^2$ is defined using the approximation of poissonian errors
as given in Ref.~\cite{Gehrels:1986mj}.  Each ensemble thus gives a $\chi^2$ 
curve
which in the region of the minimum is fitted with a fourth-order
polynomial (Fig.~\ref{fig:W_mass_hist_fit} right). The position of the
minimum of the polynomial, \mWmin, gives the best value of the W mass
and the uncertainty $\sigma(\mW)$ is obtained from the values where
$\chi^2 = \chi^2_\text{min} + 1$. The mean value of $\sigma(\mW)$
for all ensembles are quoted as expected statistical uncertainties (see Fig. 6).

\begin{figure}
  \includegraphics[width=0.35\linewidth]{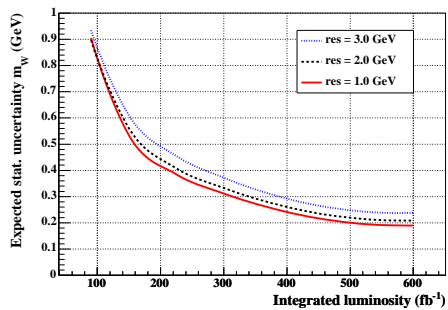}
  \caption{
    Expected statistical precision of the W mass as a function of the
    integrated luminosity for various resolutions of the roman pot
    detectors using the histogram-fitting method.}
  \label{fig:W_mass_hist_fitb}
\end{figure}

The expected statistical errors on the W mass using histogram fitting
are comparable to those using the function fitting method. However,
since the former exploits the complete missing mass distribution,
it is more sensitive to potential biases from imperfect simulation of the roman pot
detectors.

\section{Conclusion and outlook}

Recent work on DPE has essentially focused on the Higgs boson search in the exclusive channel.
In view of the difficulties and uncertainties affecting this search \cite{ourpap}, we highlight new aspects of
double diffraction which complement the diffractive program at the LHC.

In particular, QED W pair production provides a certain source of
interesting diffractive events. In this paper, we have advocated the interest of threshold scans in double
photon exchange. This method may extend the physics program at the LHC.
To illustrate its possibilities, we described in detail the \W boson 
mass measurement.
The precision of the \W mass measurement is not competitive with other methods, but provides a very precise calibration 
of the roman pot detectors, since the cross sections and characteristics of this QED process are
well under control. This method can be extended to any particle production via exclusive
processes and was applied to SUSY particle production as an example \cite{us}. 

Finally, $W$ pair production in central diffraction gives access to the coupling of gauge bosons. 
Namely, as we mentioned already, $W^+W^-$ production in two-photon exchange is robustly 
predicted within the Standard Model. Any anomalous coupling between the photon and the $W$
will reveal itself in a modification of the production cross section, or by different
angular distributions. Since the cross section of this process is proportional to the fourth]
power of the photon-$W$ coupling, a good sensitivity is expected. This study will be described
in an incoming paper \cite{usbis}.


\begin{thebibliography}{10}

\bibitem{piotr} K. Piotrzkowski, Phys. Rev. D {\bf 63} (2001) 071502.

\bibitem{Papageorgiu:1990mu}
E.~Papageorgiu,
Phys.\ Lett.\ B {\bf 250}, 155 (1990).

\bibitem{Budnev:1975zs}
V.~M.~Budnev, A.~N.~Vall and V.~V.~Serebryakov,
Yad.\ Fiz.\  {\bf 21}, 1033 (1975).

\bibitem{Binoth:2005ua}
T.~Binoth, M.~Ciccolini, N.~Kauer and M.~Kramer,
arXiv:hep-ph/0503094.

\bibitem{sp} J.~D.~Bjorken,
{Phys. Rev. D} {\bf  47}, (1993) 101; E. Gotsman, E. Levin and U. Maor,
{Phys. Lett. B}  {\bf 438} (1998), 229;
A. B. Kaidalov, V. A. Khoze, A. D. Martin and M. G. Ryskin,
{Eur. Phys. J. C}  {\bf 21} (2001) 521;
A. Bialas, `{Acta Phys. Polon. B} {\bf 33} (2002) 2635;
A. Bialas, R. Peschanski, `{Phys. Lett. B} {\bf 575} (2003) 30.




\bibitem{pom}\rr{A.~Donnachie, P.~V.~Landshoff} {Phys. Lett. B} {207}
{(1988) 319}.



\bibitem{Khoze:2001xm}
V.~A.~Khoze, A.~D.~Martin and M.~G.~Ryskin,
Eur.\ Phys.\ J.\ C {\bf 23}, 311 (2002)

\bibitem{dpemc} 
M. Boonekamp, T. Kucs, Comput. Phys. Commun. {\bf 167} (2005) 217.

\bibitem{herwig}
G. Corcella et al., JHEP {\bf 0101:010} (2001).

\bibitem{helsinki} 
J. Kalliopuska, T. M\"aki, N. Marola, R. Orava, K. \"Osterberg, M. Ottela, 
HIP-2003-11/EXP.

\bibitem{Albrow:2000na}
M.~G.~Albrow and A.~Rostovtsev [arXiv:hep-ph/0009336].


\bibitem{cmsim} 
CMSIM, fast simulation of the CMS detector, CMS Collab., Technical Design Report 
(1997);\\
TOTEM Collab., Technical Design Report, CERN/LHCC/99-7;\\
ATLFAST, fast simulation of the ATLAS detector, ATLAS Collab, Technical Design 
Report, CERN/LHC
C/99-14.

\bibitem{Abbiendi:2002ay}
  G.~Abbiendi {\it et al.}  [OPAL Collaboration],
  Eur.\ Phys.\ J.\ C {\bf 26}, 321 (2003)
  [arXiv:hep-ex/0203026].

\bibitem{Gehrels:1986mj}
  N.~Gehrels,
  Astrophys.\ J.\  {\bf 303}, 336 (1986).


\bibitem{ourpap}
M.~Boonekamp, R.~Peschanski and C.~Royon,
Nucl.\ Phys.\ B {\bf 669}, 277 (2003);
M.~Boonekamp, R.~Peschanski and C.~Royon,
Phys.\ Lett.\ B {\bf 598}, 243 (2004);
M.~Boonekamp, R.~Peschanski and C.~Royon,
Phys.\ Rev.\ Lett.\  {\bf 87}, 251806 (2001)


\bibitem{us} 
M. Boonekamp, J. Cammin, S. Lavignac, R. Peschanski, C. Royon, Phys. Rev. D {\bf 73} (2006)
115011.

\bibitem{usbis} 
M. Boonekamp, O.Kepka, R. Peschanski, C. Royon, in preparation.


\end{thebibliography}
\end{document}